\documentclass[review,12pt]{elsarticle}

 \usepackage{graphicx}
 \usepackage{epstopdf}

\usepackage{amssymb}
\usepackage[caption=false]{subfig} 

\biboptions{sort&compress}

\begin{document}

\begin{frontmatter}



\title{Polar octahedral rotations: a path to new multifunctional materials}


\author{Nicole A.\ Benedek\corref{cor1}} 
\author{Andrew T.\ Mulder\corref{cor2}}
\author{Craig J.\ Fennie\corref{cor3}}
\cortext[cor1]{nab83@cornell.edu}
\cortext[cor3]{fennie@cornell.edu}

\address{School of Applied and Engineering Physics, Cornell University, Ithaca, New York 14853 USA}

\begin{abstract}
Perovskite ABO$_3$ oxides display an amazing variety of phenomena that can be altered by subtle changes in the chemistry and internal structure, making them a favorite class of materials to explore the rational design of novel properties. Here we highlight a recent advance in which rotations of the BO$_6$ octahedra give rise to a novel form of ferroelectricity -- hybrid improper ferroelectricity. Octahedral rotations also strongly influence other structural, magnetic, orbital, and electronic degrees of freedom in perovskites and related materials. Octahedral rotation-driven ferroelectricity consequently has the potential to robustly control emergent phenomena with an applied electric field.  The concept of `functional' octahedral rotations is introduced and the challenges for materials chemistry and the possibilities for new rotation-driven phenomena in multifunctional materials are explored.
\end{abstract}


\end{frontmatter}


\section{Introduction}
\begin{figure}
\centering
  \includegraphics[height=5cm]{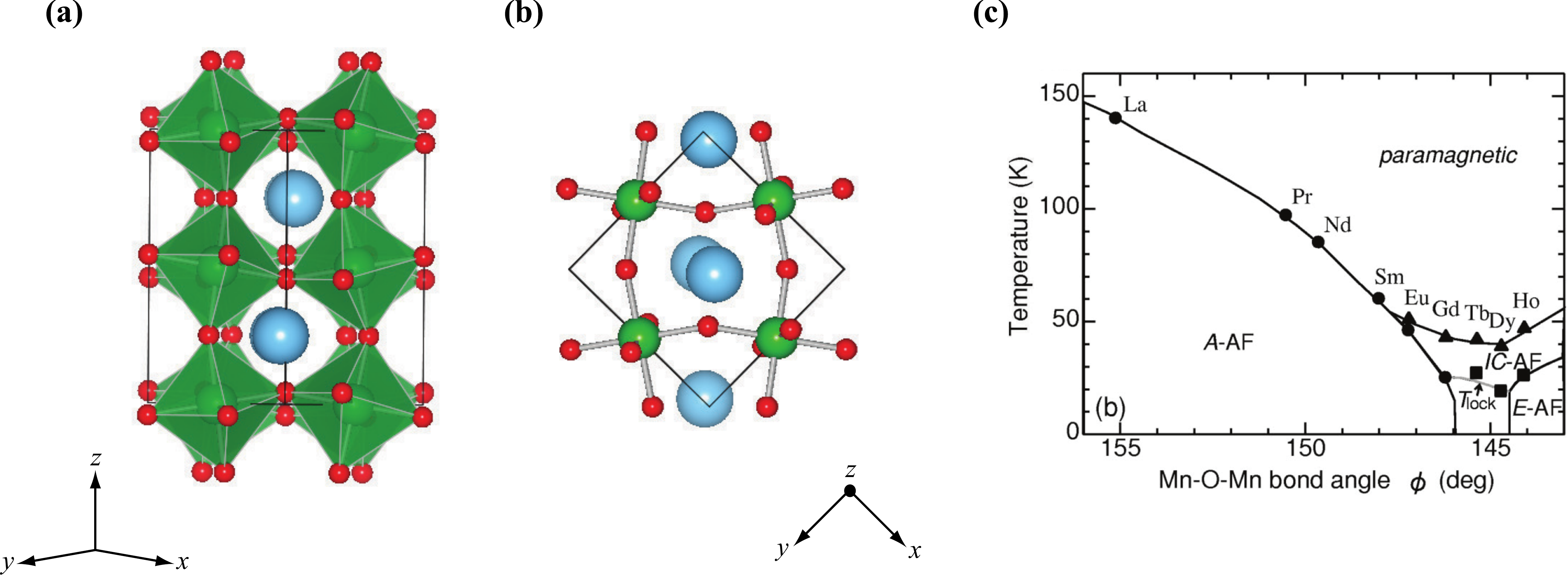}
  \caption{(a) Nonpolar $Pnma$ structure adopted by many perovskites. (b) Alternative view showing the bending of the TM-O-TM bonds. (c) Change in magnetic properties and ordering temperatures as a function of Mn-O-Mn bond angle for $Pnma$ AMnO$_3$ perovskites, A = La$\rightarrow$Ho. From Ref. \cite{kimura03b}.}
  \label{pnma}
\end{figure}

Most ABO$_3$ perovskites (and related materials) undergo structural distortions associated with a `rotation' of the BO$_6$ octahedra about one or more of the crystal axes.\cite{glazer72,woodward97a,woodward97b,mitchell02} The orthorhombic $Pnma$ structure of the mineral perovskite (CaTiO$_3$) is a combination of two octahedral rotations about different crystallographic axes [Figure \ref{pnma}(a)]; the majority of perovskites adopt this same structure. 
The B site in ABO$_3$ perovskites is commonly occupied by a transition metal and octahedral rotations can significantly change the transition metal-oxygen-transition metal (TM-O-TM) bond angles, as shown in Figure \ref{pnma}(b). 
Consider the case of magnetism in perovskite AMnO$_3$ manganites, where A is typically an alkaline-earth and/or a rare-earth cation(s).  Through substitution on the A-site with cations of different radii it is possible to control the degree to which the BO$_6$ octahedra are rotated and hence the TM-O-TM bond angle. Such changes in bond angles can have a profound effect on the interaction between spins. 
As an example, in the simplest case of spins interacting through superexchange, materials in which the TM-O-TM  bond angles are closer to 180$^{\circ}$ are generally antiferromagnetic, whereas materials in which the TM-O-TM bond angles are closer to 90$^{\circ}$ tend to be ferromagnetic\cite{goodenough55, kanamori57,anderson59}. 
Specifically in the rare-earth manganites, small changes in the Mn-O-Mn angle induced via cation substitution on the A-site change the magnetic ground state from collinear A-type antiferromagnetism (AFM), to a spin-spiral, to E-type AFM,\cite{kimura03b} as shown in Figure \ref{pnma}(c). Octahedral rotations consequently have the potential to robustly control the magnetic properties, not to mention the electronic, orbital, and dielectric properties, of a given material~\cite{goodenough55,millis98,goto04,rini07,lufaso04,rondinelli10}.

The strong coupling of oxygen rotations to properties represents an opportunity to understand and create new functional materials that respond to an external perturbation in a useful way. For example, one of the biggest challenges in multiferroics research is finding or designing so-called \emph{cross-coupled} multiferroics. These are materials that are simultaneously magnetic and ferroelectric but the polarization (magnetization) can be manipulated in a useful way with an applied magnetic (electric) field. We know that octahedral rotations can significantly affect the magnetic properties of perovskites, but rotations (or combinations of rotations) in simple perovskites do not directly couple to an external field (other than pressure\cite{samara75} or stress\cite{rond_spaldin11}). In this article, we review progress on a recent development in which the layering of nominally nonpolar perovskites leads to a new type of ferroelectricity in which octahedral rotations induce an electrical polarization. Electric fields couple naturally to the polarization in an insulator and therefore, in these rotation-driven ferroelectrics, an applied electric field can directly couple to the octahedral rotations and can even switch the sense of the rotations.

This article is organized as follows. In Section \ref{origin_design} we discuss the origin of and design rules for ferroelectric perovskites and perovskites with octahedral rotations. We also define and explain the differences between proper and improper ferroelectrics. We introduce the concept of hybrid improper ferroelectricity in Section \ref{hif} and present several examples of hybrid improper ferroelectrics from the recent literature. The crystal chemical origin of this new type of ferroelectricity is elucidated and design rules for the creation of new materials are presented. Finally, in Sections \ref{coupling} and \ref{end}, we speculate on the manner in which hybrid improper ferroelectricity could be exploited to create new multifunctional materials.

Before we begin, we should point out that the ideas that we discuss, the generalizations and commentary we make, refer almost exclusively to perovskite and perovskite-like {\it oxides}.
Additionally, when we speak of ``ferroelectricity'' we are referring to polar structures in which the polarization can in principle be switched to a symmetry-equivalent state with an applied electric field and that satisfy the simple structural criteria devised by Abrahams and co-workers.\cite{abrahams68}



%
%
\section{\label{origin_design}Octahedral Rotations and Ferroelectricity: Origin and Design Rules}
One strategy to couple octahedral rotations to an electric field is to start with a ferroelectric material in which the polarization is induced by rotations. A materials chemistry challenge is understanding how to rationally design and synthesize such materials. As a starting point, let us first recall the origin of and design rules for prototypical perovskite ferroelectrics and for perovskites with rotations of the BO$_6$ octahedra. We will also introduce the notion of an improper ferroelectric, which we will revisit in a later section.

%
%
\subsection{Ferroelectric Mechanisms: Proper and Improper}
\begin{figure}
\centering
  \includegraphics[height=3.0cm]{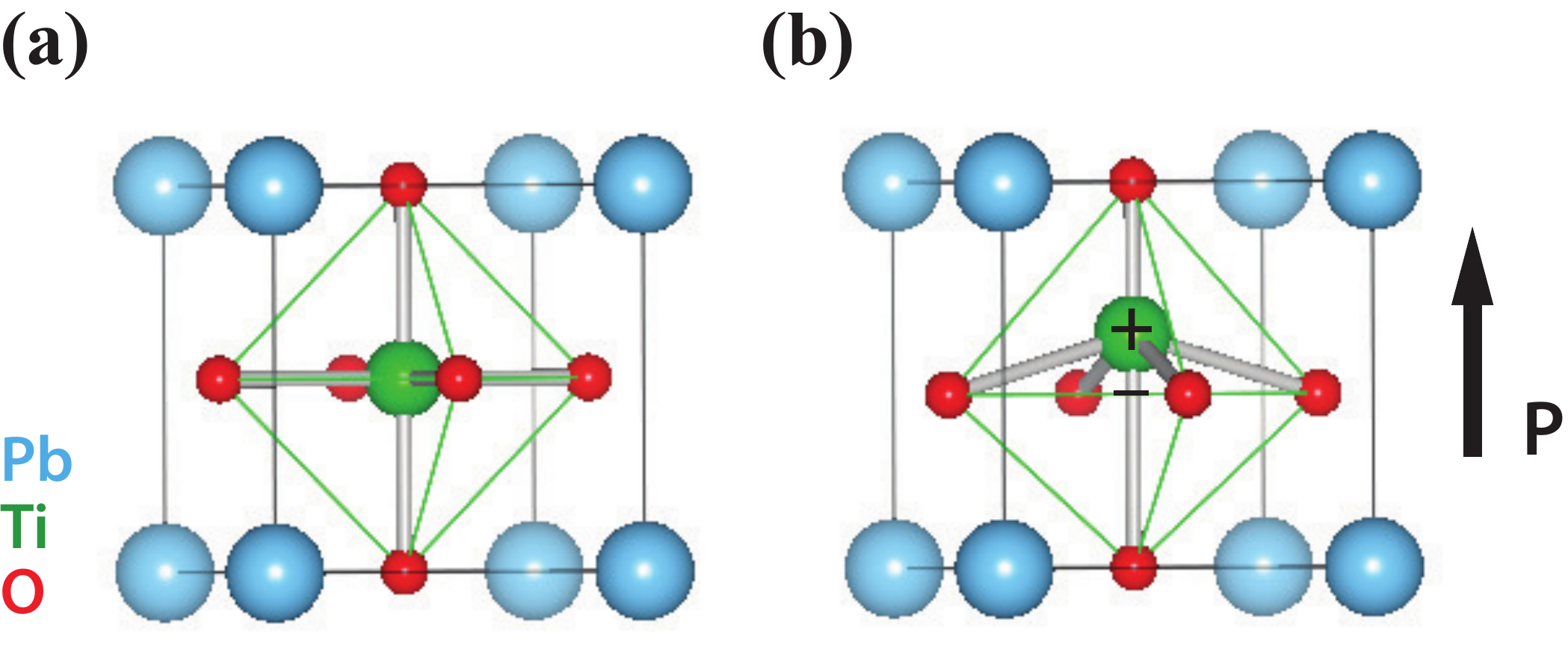}
  \caption{(a) Paraelectric $Pm\bar{3}m$ structure of PbTiO$_3$. (b) Ground state ferroelectric $P4mm$ structure of PbTiO$_3$. A displacement of the Ti and Pb cations against the oxygen anions gives rise to a spontaneous polarization and relates the ground state to the nonpolar $Pm\bar{3}m$ structure that PbTiO$_3$ is found in above $\sim$760 K. The cell also undergoes a spontaneous strain from cubic to tetragonal.}
  \label{FE}
\end{figure}
{\it Proper} -- PbTiO$_3$ is perhaps the archetypal ferroelectric perovskite and its ground state (lowest energy) structure is shown in Figure \ref{FE}(b). As we explain below, prototypical perovskite ferroelectrics, such as PbTiO$_3$, display what crystallographers would call a proper ferroelectric transition: a polar lattice distortion has a negative force constant and its condensation completely accounts for the difference in symmetry between the paraelectric parent phase and the ferroelectric ground state.\cite{perez-mato10,cohen92,cochran59}

Cohen showed\cite{cohen92} that the origin of ferroelectricity in PbTiO$_3$ and BaTiO$_3$ is hybridization between the formally empty Ti 3$d$ states and filled O 2$p$ states. This mechanism is described within the framework of vibronic coupling theory\cite{bersuker01,halasyamani98} as a second-order Jahn-Teller effect and in this sense is formally similar to the origin of ferroelectricity in lone-pair compounds\cite{singh05, payne06} such as BiFeO$_3$\cite{seshadri01}.
\begin{figure}
\centering
  \includegraphics[height=3.5cm]{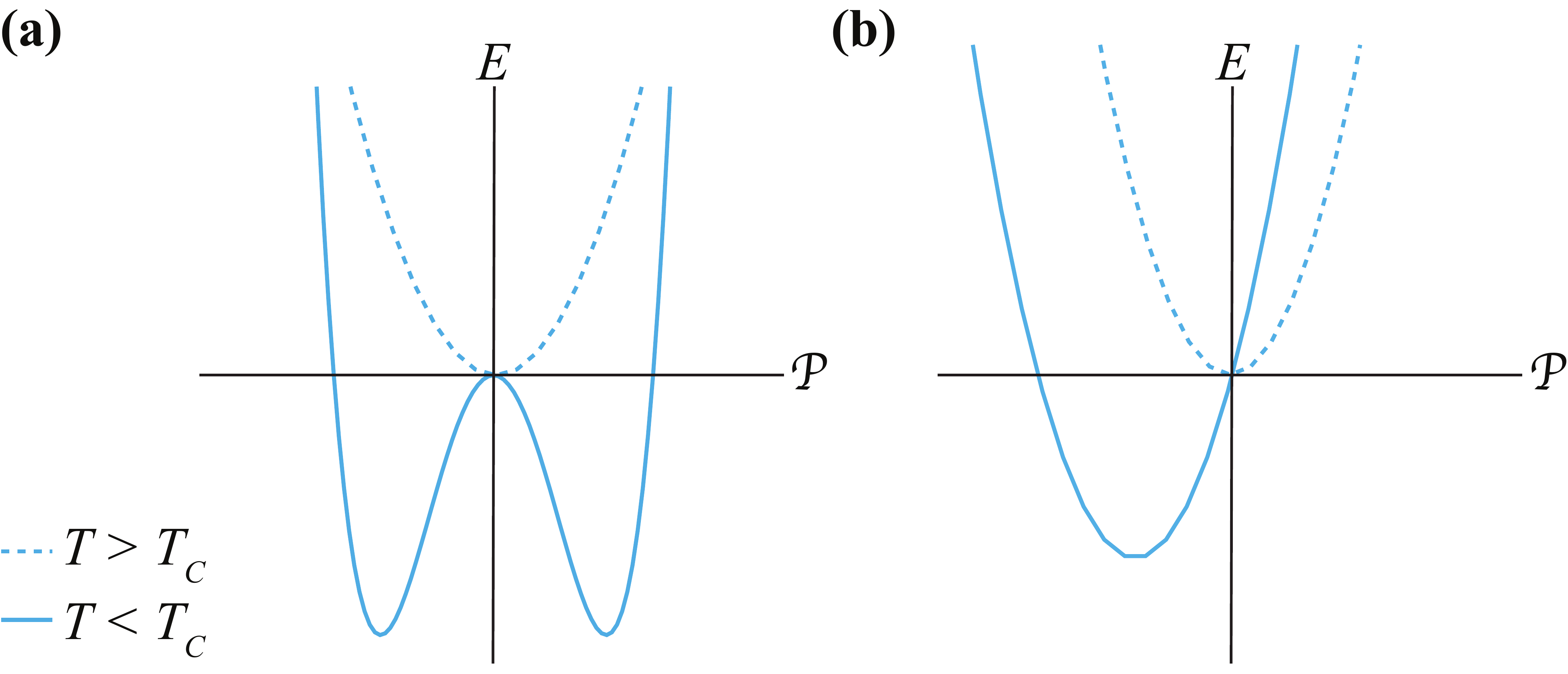}
  \caption{The behavior of the energy as a function of the polarization when the temperature, $T$, is above and below the Curie temperature ($T_C$) of the primary order parameter for a) a proper ferroelectric transition and b) an improper ferroelectric transition.}
  \label{transitions}
\end{figure}

In a complementary picture, ferroelectricity arises from the `freezing-in' or condensation of a polar lattice distortion $u_{\bf q}$, that is, a pattern of atomic displacements that breaks all symmetries (including inversion symmetry) that would otherwise forbid the appearance of ferroelectricity. These distortions have negative force constants and arise at the Brillouin zone center ($\mathbf{q}=0$, where $\mathbf{q}$ is the wave vector of the lattice distortion). 

In proper ferroelectrics, the polarization is proportional to the amplitude of the polar lattice distortion ($u_{\bf q =0}\propto P$) and the energy around the paraelectric structure displays a characteristic double-well potential described by the equation,
\begin{equation}
\mathcal{F}_P =  \xi_P\,P^2 + \zeta_P\,P^4,
\label{quadratic}
\end{equation}
where $\mathcal{F}_P$ is the free energy. As shown in Figure \ref{transitions}(a), $\xi$ is less (greater) than zero in the ferroelectric (paraelectric) phase and reflects the force constant instability (stability). In principle, an electric field can be used to switch the polarization between the two minima without any other structural distortions being involved. 
%

\begin{figure}
\centering
\includegraphics[height=7.5cm]{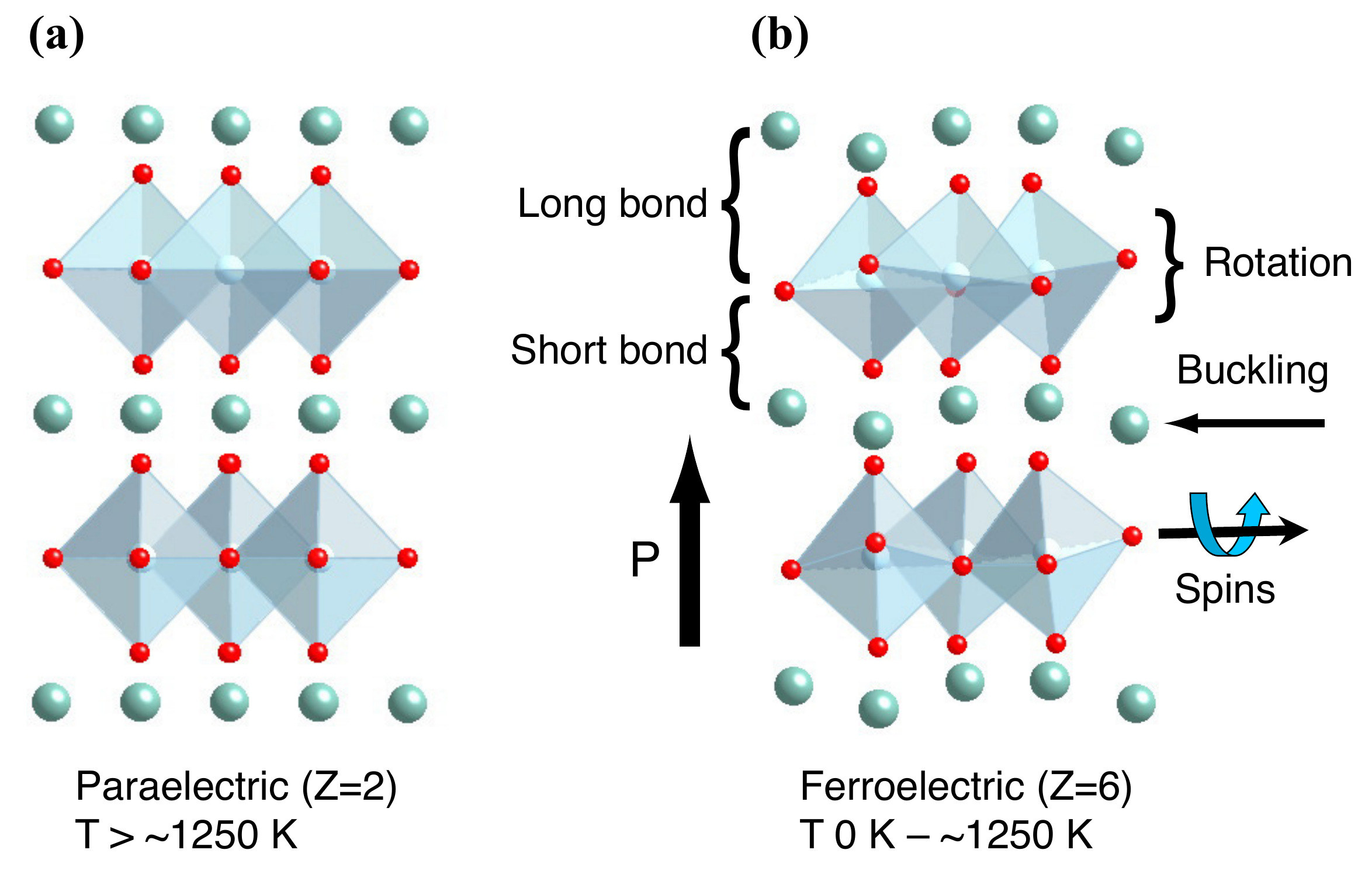}
\caption{(a) Structure of the paraelectric phase of the improper ferroelectric YMnO$_3$. (b) Structure of the ferroelectric phase and direction of the polarization, $P$.}
\label{YMnO3}
\end{figure}
{\it Improper} -- Unlike the proper ferroelectrics we just discussed, prototypical improper ferroelectrics~\cite{levanyuk74, cheong07} have no zone-center polar instabilities. Instead, the polarization is a `slave' to some other primary structural distortion associated with a wave vector at the zone-boundary. Hence, large structural changes occur at the paraelectric to ferroelectric transition, \emph{e.g.,} the unit cell volume increases by at least a factor of two. 
In the case of a large number of hexagonal manganites, such as YMnO$_3$~\cite{fennie05,vanaken04,jeong07,lightfoot11}, the primary distortion is  a single unstable mode with $\mathbf{q} = (\frac{1}{3},\frac{1}{3},0)$ (the $K$ mode) consisting of a rotation of the MnO$_5$ trigonal-bipyramids and a buckling of the yttrium planes, which results in a tripling of the unit cell volume (Figure~\ref{YMnO3}).

At the phenomenological level, this mechanism can be understood~\cite{fennie05} as arising from a term in the free energy of the form 
\begin{equation}
\mathcal{F}_{1f} = \alpha\,P\,K^f
\label{improper}
\end{equation}
(with $f$=3 for the hexagonal manganites by symmetry) where, because of the symmetry-allowed coupling between $P$ and $K$, once $K$ becomes finite, a non-zero value of $P$ is induced. However, unlike in a proper ferroelectric, in an improper ferroelectric the polarization remains in a single-well minimum, with the minimum shifted to a non-zero value, as shown in Figure \ref{transitions}(b). This is analogous to turning on an electric field, $E$, 
\begin{equation}
\mathcal{F}_{1E} = -P\cdot E. 
\end{equation}
To switch the polarization in an improper ferroelectric, the primary distortion (the oxygen rotation-like mode, in the case of YMnO$_3$) must also switch. In other words, if $K$ were an octahedral rotation, the sense of the rotation must change if the sign of $P$ is reversed.

YMnO$_3$ is not a perovskite, so even though an electric field would switch the sense of the rotation-like distortion of the MnO$_5$ trigonal bipyramids, this distortion does not affect the magnetic properties in any meaningful way. Furthermore, although we understand -- from a symmetry perspective -- the general mechanism through which the polarization couples to the primary order parameter in YMnO$_3$, we have no general rules for designing improper ferroelectrics. An interesting question is whether or not we can design a perovskite improper ferroelectric in which octahedral rotations are the primary order parameter. In such a material, the polarization would be completely accounted for by the octahedral rotations.


%
%
%
\subsection{Octahedral Rotations}
Octahedral rotations are associated with a lattice instability, $u_{\bf q}$, of the force constant matrix at the Brillouin zone boundary of the cubic unit cell and so always involve an increase in the unit cell volume.
In this case, atomic displacements in neighboring unit cells are related by a phase factor that alternates sign along certain directions from one unit cell to the next. This varying phase factor cancels any local dipoles, maintains the octahedral corner-connectivity and therefore results in zero macroscopic polarization. In other words, because of the three-dimensional connectivity of the octahedra in the perovskite structure, octahedral rotations cannot by themselves induce ferroelectricity. This was proved by Stokes and co-workers\cite{stokes02} using group theoretical techniques.

In prototypical perovskites with octahedral rotations (such as SrTiO$_3$) the rotation pattern, $R$, is proportional to the amplitude of $u_{\bf q}$ and the ground state displays a characteristic double-well potential described by
\begin{equation}
\mathcal{F}_R =  \xi_R\,R^2 + \zeta_R\,R^4,
\label{rotation_Energy}
\end{equation}
where again $\xi < 0$ implies a distorted ground state. Prototypical perovskites with octahedral rotations display a proper antiferrodistortive transition where, as was the case for proper ferroelectric transitions, the rotation lattice distortion has a negative force constant and completely accounts for the symmetry lost between the high-symmetry parent phase and the ground-state structure with octahedral rotations.

A well-understood, albeit heuristic, approach to control and design ABO$_3$ perovskites that display octahedral rotations involves the idea of the tolerance factor, 
\begin{equation}
\tau = {R_O+R_A\over \sqrt{2} (R_O+R_B),}
\end{equation}
a geometric measure of closed-sphere packing relating A-O and B-O bond lengths to crystal stability, where $R_O$, $R_A$, and $R_B$ are the radii of the oxygen, A-site cation and B-site cation respectively. 
Compounds with tolerance factors between roughly $0.9<\tau<1.0$ form with octahedral rotations and are usually not ferroelectric (unless they contain lone-pair active ions such as Bi$^{3+}$ or Pb$^{2+}$ on the A-site).\footnote{It is interesting to note that an increasing number of first-principles studies of the lattice dynamics of perovskites show that both rotation and ferroelectric instabilities coexist in the energy landscapes of many materials\cite{kingsmith94,ghosez99,cockayne00,bilc06}, \textit{e.g.}, CaTiO$_3$.\cite{parlinski,cockayne00,rabe09} However, whereas there are many perovskites that are \emph{either} ferroelectric or have rotated octahedra, there are few perovskites (again, without lone-pair active ions) that are \textit{both} ferroelectric and have rotated octahedra.}
Changing the tolerance factor of a compound can be experimentally accomplished by either applying external pressure or substitution doping of the A or B sites (or both), where the substituted cation has a different ionic radius (this is akin to applying an internal or `chemical' pressure).

Perovskites that display proper ferroelectricity or proper antiferrodistortive transitions can be designed by a suitable choice of chemistry, \emph{e.g.,} Jahn-Teller ions or A/B ion size mismatch. However, group theory and structural analysis tells us that the type of structural distortion associated with polar distortions in proper ferroelectrics tends to occur independently of rotations of the octahedra.

%
%
\section{\label{hif}Hybrid Improper Ferroelectricity}

Although octahedral rotations cannot by themselves induce ferroelectricity in simple perovskites, it has recently been (re)appreciated that the \textit{combination} of octahedral rotations and layered cation ordering can in fact give rise to ferroelectricity in perovskite oxides\cite{bousquet08,fukushima11,jorge11,rondinelli11,benedek11}. This was anticipated from the earlier work of Woodward and co-workers\cite{knapp06,woodward07,king10}. 
In this section, we discuss the discovery of octahedral rotation-induced ferroelectricity in several different materials. We introduce the concept of hybrid improper ferroelectricity to describe this type of ferroelectricity, describe its structural origin and present design rules for identifying and designing new hybrid improper ferroelectrics. 
%

\subsection{(SrTiO$_3$)$_n$/(PbTiO$_3$)$_m$ superlattices}
In a recent combined theoretical-experimental study, Bousquet, Dawber and co-workers grew (SrTiO$_3$)$_n$/(PbTiO$_3$)$_m$ superlattices as thin-films~\cite{bousquet08}.  They found that for large period superlattices (large $n$ and $m$), as the number of PbTiO$_3$ units was reduced and, simultaneously, the number of SrTiO$_3$ units increased, the ferroelectric properties -- T$_C$ and polarization -- decreased as expected (PbTiO$_3$ being a ``good'' ferroelectric and SrTiO$_3$ being paraelectric in bulk). However, for very short period superlattices, the spontaneous polarization started to increase.
Theoretical calculations on superlattices in which one unit cell of SrTiO$_3$ alternates with one unit cell of PbTiO$_3$ revealed something quite unexpected: the polarization had an additional contribution that could be accounted for by a combination of two different octahedral rotations, $R_1$ and $R_2$, with different symmetries. In contrast to prototypical ferroelectrics like PbTiO$_3$, in the SrTiO$_3$/PbTiO$_3$ superlattices the octahedral rotations play a key role in the structure of the observed ferroelectric ground state. In analogy to YMnO$_3$, Bousquet \textit{et al.}, referred to this rotation component of the polarization as {\it improper ferroelectricity}~\cite{levanyuk74}.

%
%

In the case of the SrTiO$_3$/PbTiO$_3$ superlattices, it was possible to prove using group theoretical techniques that the combined rotation pattern of $R_1$ plus $R_2$  couples linearly to the polarization, that is, there is a trilinear term in the free energy of the form\cite{etxebarria10}
\begin{equation}
\mathcal{F}_{111} = \gamma\,P\,R_1\,R_2.
\label{trilinear}
\end{equation}
Hence, even though each rotation individually leads to a nonpolar space group, the combined rotation pattern breaks all the relevant symmetries that allow a spontaneous polarization to arise. Comparison of Eq.~\ref{improper} to Eq.~\ref{trilinear} and the experimental evidence all lead Bousquet and co-workers to refer to the SrTiO$_3$/PbTiO$_3$ superlattice as an improper ferroelectric. 

It is important to note that the trilinear term is not required to `switch on' $P$, $R_1$ and $R_2$, but it may in some cases\cite{etxebarria10}. That is, in most cases the two rotation distortions are both individually unstable, so the $\beta R_1^2$ and $\delta R_2^2$ terms in the free energy have $\beta$ and $\delta$ negative at $T=0$. However, the combined rotation pattern, $R_1$ plus $R_2$, lowers the energy more than either rotation individually. The trilinear term merely ensures in this case that the coupling of the rotations to the polarization lowers the energy even further. Because of this Benedek and Fennie\cite{benedek11} suggested calling such materials \textit{hybrid} improper ferroelectrics. 
In hybrid improper ferroelectrics like SrTiO$_3$/PbTiO$_3$, the terms that couple to a linear power of the polarization are from two different irreducible representations, \textit{i.e.}, the two rotation modes have different symmetry\cite{ghosez11}. The combined rotation pattern, $R_1$ plus $R_2$, is treated as a hybrid mode. Hybrid improper ferroelectrics may display different phase transition behavior than conventional improper ferroelectrics. The primary and secondary distortions may condense at the same temperature (in a so-called avalanche transition\cite{etxebarria10}), or there may be a series of phase transitions, where at one temperature one of the rotations condenses and then at a second temperature the second rotation condenses (technically, this second phase transition would be a proper ferroelectric transition; the work of Tagantsev on `weak' ferroelectrics most likely describes this type of transition\cite{tagantsev87}). Further experimental and theoretical investigations are currently underway to clarify this important issue. 

Equation \ref{trilinear} establishes the symmetry requirements of hybrid improper ferroelectrics, but it does not tell us about the structural or crystal chemical basis of the phenomenon. Such knowledge is crucial if we are to design new hybrid improper ferroelectrics. However, Equation \ref{trilinear} does give us a clue as to the types of materials that may display hybrid improper ferroelectricity, namely, materials whose structures consist of two different octahedral rotation patterns. 

%
%
%
\subsection{Pnma Perovskites as `Building Blocks' to Create New Hybrid Improper Ferroelectrics}
\begin{figure}
\centering
\includegraphics[width=11cm]{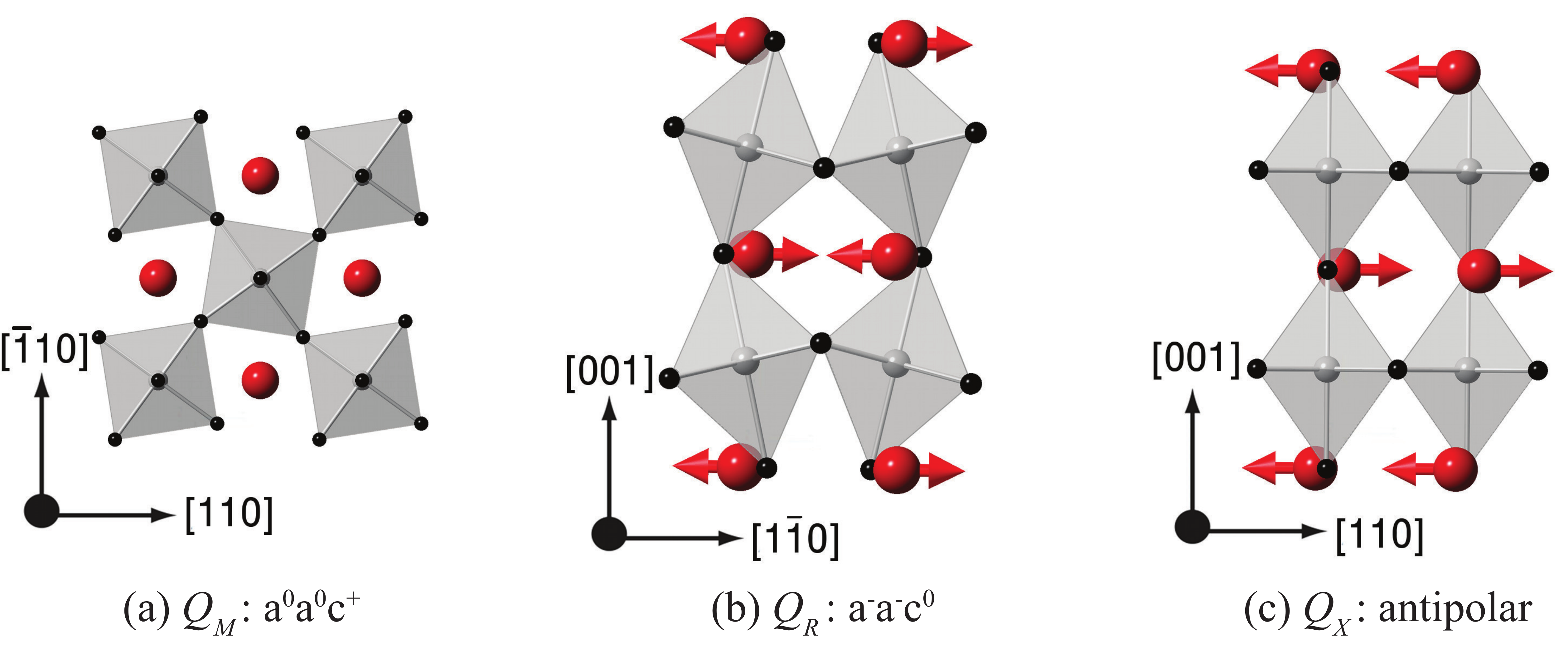}
\caption{The \textit{Pnma} structure (rotation pattern a$^-$a$^-$c$^+$) can be described primarily by three different normal modes of \textit{Pm$\bar{3}$m}: (a) Octahedral rotation about [001] from a mode at the M-point, (b) rotation about [110] with A-site displacement from a mode at the R-point, and (c) antipolar A-site displacement without rotation from a mode at the X-point. 
}
\label{ABO3}
\end{figure}

Recently, Rondinelli and Fennie\cite{rondinelli11} identified the \textit{Pnma} structure as an important building block in the design of hybrid improper ferroelectrics. This was a surprising and counterintuitive result because the $Pnma$ space group is nonpolar. The symmetry of the \textit{Pnma} structure is established by two octahedral rotations, shown in Figure \ref{ABO3}(a) and (b): $Q_M$ (transforming like $M_3^+$, with a rotation pattern a$^0$a$^0$c$^+$ in Glazer notation) and $Q_R$ (transforming like $R_4^+$, with a$^-$a$^-$c$^0$). The \textit{Pnma} structure is the most common rotation pattern among ABO$_3$ perovskites,~\cite{woodward97a,woodward97b,lufaso01} which means that there are many materials that could potentially be used to create new hybrid improper ferroelectrics. How can we use $Pnma$ perovskites to create these new materials?
%

\subsubsection{The Importance of A-site Cation Displacements in the Pnma Structure}
\begin{figure}
\centering
\includegraphics[height=10cm]{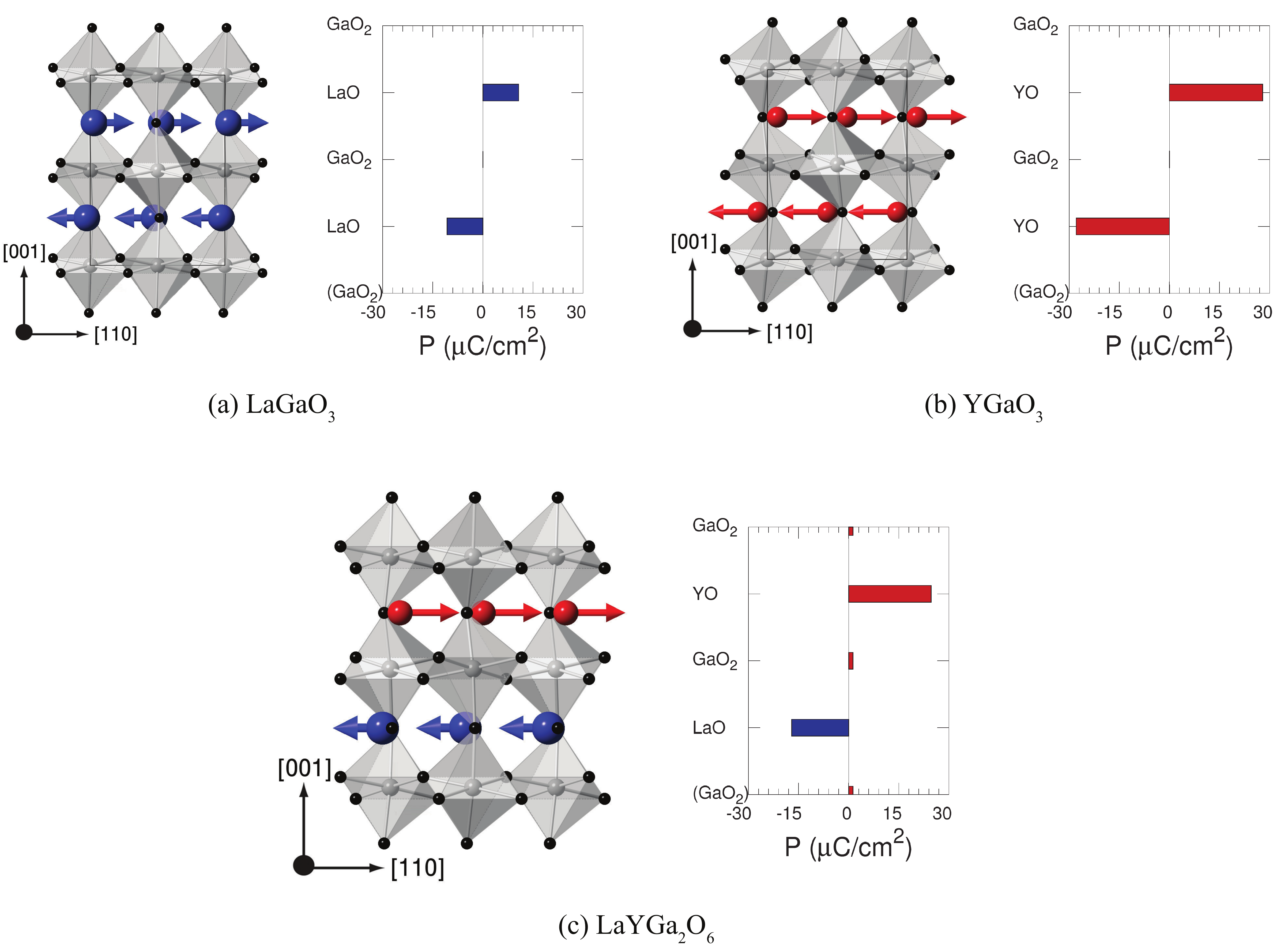}
\caption{The layer-resolved polarization in the \textit{Pnma} ground state of (a) LaGaO$_3$ and (b) YGaO$_3$. To the right of each structure is shown a plot of the layer-resolved polarization, $P_{\mathrm{layer}} = \frac{1}{\Omega}\sum_i \mathbf{u}_i Z^\ast_{i}$, where $\Omega$ is the cell volume, $\mathbf{u}_i$ is the displacement of ion $i$ from its position in a higher-symmetry reference structure ($Pmma$ for the superlattice and $Imma$ for the ABO$_3$ perovskites) and $Z^\ast_{i}$ is the Born effective charge of ion $i$. The total polarization is the sum of the layer polarizations, $P = \sum P_{\mathrm{layer}}$. The axes shown refer to the the cubic \textit{Pm$\bar{3}$m} cell. Symmetry-enforced cancellation of the dipole moments leads to $P = 0$. (c) In the LaYGa$_2$O$_6$ superlattice, inequivalent A-site displacements lead to imperfect cancellation of the layer polarizations. The polarization is parallel to [110]. \textit{Methods: VASP, PAW-PBE potentials, PBEsol functional, \textbf{k}-grid equivalent to 8$\times$8$\times$8 in the cubic cell, 600eV cutoff. Born effective charges calculated using DFPT in the ground state structure ($Pmc2_1$ for the superlattice, $Pnma$ for the ABO$_3$ perovskites).}}
\label{P_Ga}
\end{figure}

Octahedral rotations in perovskites are driven by the coordination preferences of the A-site cation~\cite{woodward97a,woodward97b}. In some space groups with octahedral rotations, the A-site cation can shift from its `ideal' $Pm\bar{3}m$ position to further optimize its coordination environment. 
In $Pnma$, the A-site cation can displace in an antipolar pattern. This pattern can be thought of as a ferroelectric-like displacement of the A-site in two-dimensional layers arranged 180$^{\circ}$ out of phase along $\hat{z}$, as shown in Figure~\ref{ABO3}(c). We denote this distortion as $Q_X$, since it transforms like the irreducible representation $X_5^+$. The displacement of the A-site cations in $Pnma$ induces a significant polarization within the AO layers. This can be seen from first-principles calculations in which the polarization is resolved into AO and BO$_2$ layers; Figures \ref{P_Ga}(a) and \ref{P_Ga}(b) show the results of such a calculation for the $Pnma$ perovskites LaGaO$_3$ and YGaO$_3$.

\begin{figure*}
\includegraphics[width=0.95\textwidth]{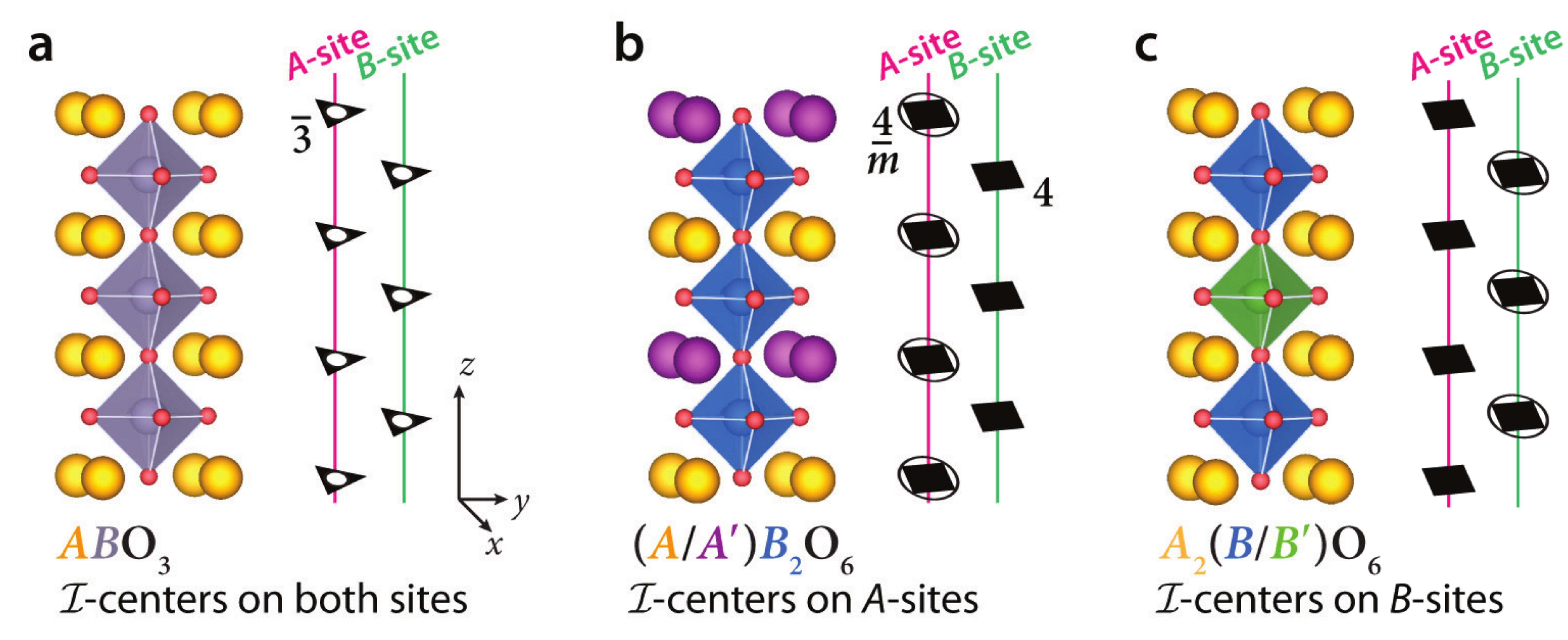}\vspace{-7pt} 
\caption{\label{fig:ingredients} Illustration of the chemical criterion for 
rotation-induced ferroelectricity in layered perovskite superlattices constructed from two different $AB$O$_3$ perovskite materials.
In (a) bulk $AB$O$_3$ perovskites inversion ($\mathcal{I}$) centers are found on both the $A$- and $B$-sites; the highest site-symmetry operator being a three-fold rotoinversion ($\bar{3})$.
Cation ordering in layered perovskites, however, lifts the inversion centers on
the $B$-site (leading to a 4-fold rotation) in the $A/A^\prime$ layered perovskites (b) and on the $A$-site in the $B/B^\prime$  (c) structures.
Inversion only remains through the $\frac{4}{m}$ operation found on the remaining $A$-site  and $B$-site, respectively. 
Since rotations of octahedra preserve the inversion on $B$-sites yet can remove it on the $A$-sites, only $A/A^\prime$ 
support this form of hybrid improper ferroelectricity. From Ref. \cite{rondinelli11}.
}
\end{figure*}

What is the importance of A-site cation displacements to hybrid improper ferroelectricity? In the cubic perovskite \textit{Pm$\bar{3}$m} reference structure, both the A and B sites lie on an inversion center and three mirror planes perpendicular to the coordinate axes, as shown in Figure~\ref{fig:ingredients}(a). Because of the three-dimensional connectivity of the octahedra in the perovskite, $Q_M$ and $Q_R$ together preserve the inversion center $\mathcal{I}$ at each B-site (in fact, octahedral rotations can never by themselves remove the inversion center at the B-site, the consequences of which will be made clear in the next section), and therefore the total polarization must be zero by symmetry. However, $Q_M$ and $Q_R$ break the mirror planes that would otherwise forbid a net displacement of the A-cations against the oxygens in each AO layer. Hence, although the total polarization must be zero in the \textit{Pnma} structure, the amplitude of $Q_X$ can be significant.  
Note that the antipolar A-site displacements need not be unstable to appear in the $Pnma$ structure. Indeed, most perovskites, including LaGaO$_3$ and YGaO$_3$, do not favor this motion in the absence of the $Pnma$ a$^-$a$^-$c$^+$ rotation pattern. This can easily be seen from first-principles calculations of the force constants within $Pm\bar{3}m$ (the  $X_5^+$ phonon is stable).

The appearance of a finite $Q_X$ in the $Pnma$ structure can be accounted for phenomenologically by a trilinear coupling in the free energy of $Pm\bar{3}m$, $\mathcal{F}_{\rm tri} \sim Q_M Q_R Q_X$. This invariant couples the $Pnma$ rotations, $Q_M$ and $Q_R$, to the antipolar motion of the A-site cations, $Q_X$, ensuring that once $Q_M$ and $Q_R$ become nonzero, a finite $Q_X$ is induced\cite{ghosez12} even when stable in $Pm\bar{3}m$. 

\begin{figure}
\centering
\includegraphics[height=4.5cm]{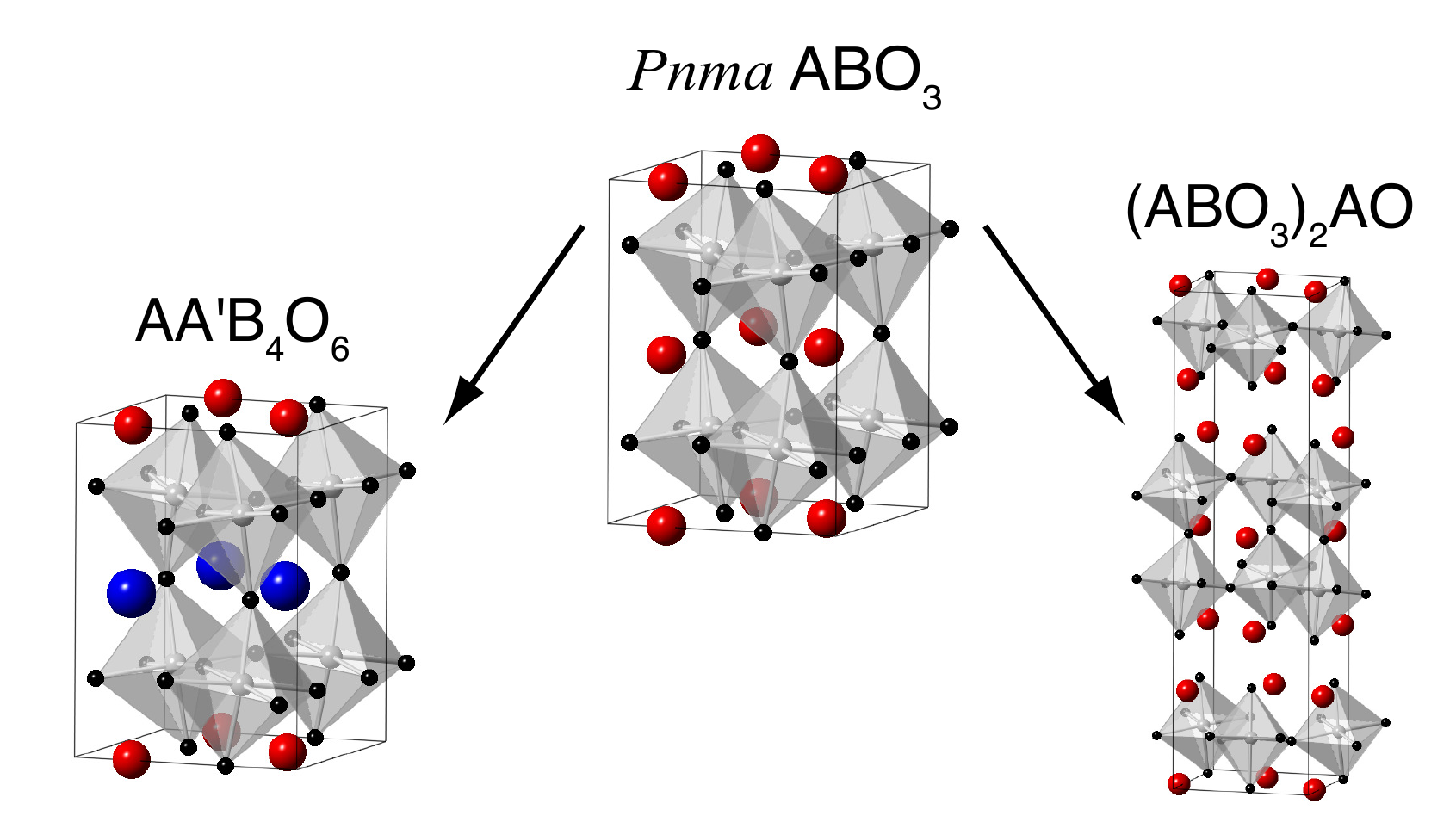}
\caption{Two alternative structure types that can be created by layering $Pnma$ ABO$_3$ perovskites in different ways: A-site ordered double perovskites (left) and Ruddlesden-Popper phases (right). Hybrid improper ferroelectricity is allowed by symmetry in both of the layered structures shown.}
\label{Pnma_SL_RP}
\end{figure}

This trilinear coupling is reminiscent of that found in hybrid improper ferroelectrics, Eq.~\ref{trilinear}. Is it related?  Any perturbation to the $Pnma$ structure that breaks the inversion center at each B-site would lead to inequivalent A-site layers, and hence to a small noncancellation of the polarization induced by $Q_X$ in the AO layers. But since rotations alone cannot remove this symmetry, some other ``distortion'' is required to create inequivalent A-site environments.  As we will now discuss, this non-cancellation, or {\it ferri-electric} mechanism, is the origin of hybrid improper ferroelectricity in materials built from $Pnma$ perovskites. Figure \ref{Pnma_SL_RP} shows two structural motifs that allow for the existence of hybrid improper ferroelectricity: A-site ordered double perovskites and Ruddlesden-Popper phases.

%
%
\subsubsection{Hybrid Improper Ferroelectricity in AA$^\prime$B$_2$O$_6$ Double Perovskites}

Perhaps the simplest way to create a perovskite with inequivalent A-sites is to replace alternating AO layers with A$^\prime$O to form a superlattice with formula AA$^\prime$B$_2$O$_6$. This cation ordering strategy for rotation-driven ferroelectricity was recently discussed by Rondinelli and Fennie~\cite{rondinelli11} who have demonstrated how ferroelectric structures can be rationally designed from nonpolar $Pnma$ building blocks. 
Specifically, using group-theoretical methods combined with \textit{ab initio} density functional theory calculations on gallate and aluminate-based ABO$_3$ perovskites, they outlined a universal set of material structure-property relationships for realizing hybrid improper ferroelectricity in perovskites with AA$^\prime$B$_2$O$_6$ stoichiometry.
The guidelines were separated into an energetic criterion and a chemical criterion. The energetic criterion requires that the bulk perovskite of one or more of the constituents of the superlattice have a strong tendency towards the $Pnma$ structure (that is, $Pnma$ should be the ground state structure, preferably, or a metastable phase with a wide stability window). This ensures that the a$^-$a$^-$c$^+$ rotation pattern survives in the superlattice. The chemical criterion, summarized in Figure~\ref{fig:ingredients}, recognizes that Glazer rotations in perovskites cannot break the inversion center on the B-site, therefore some form of cation ordering, \textit{e.g.,} A/A$^\prime$ layered ordering, is required.
A particularly elegant aspect of the design criteria is that they relate the properties of the AA$^\prime$B$_2$O$_6$ superlattice back to the properties of the parent single-phase ABO$_3$ and A$^\prime$BO$_3$ perovskites. Hence, at least in principle, it should be possible to predict whether a particular AA$^\prime$B$_2$O$_6$ material will be a hybrid improper ferroelectric just by considering the properties of the constituent ABO$_3$ and A$^\prime$BO$_3$ phases.   

As a specific example of creating a hybrid improper ferroelectric according to the guidelines of Rondinelli and Fennie, one could replace alternating LaO layers in LaGaO$_3$ with YO to form a LaYGa$_2$O$_6$ superlattice, as shown in Figure \ref{P_Ga}(c). 
Given the strong tendency of both LaGaO$_3$ and YGaO$_3$ to form in the $Pnma$ structure, the superlattice displays the same a$^-$a$^-$c$^+$ rotation pattern. As we have discussed, these rotations favor an antipolar displacement of the La and Y ions in the LaO and YO layers and the lack of an inversion center on the B-site means that the polarization induced by these displacements does not exactly cancel, leading to a net polarization [Figure \ref{P_Ga}(c)]. Thus, the rotations which drive the system to the polar space group, \textit{Pmc2$_1$}, induce a net ferri-electric polarization.

\begin{figure}
\centering
\includegraphics[height=8.0cm]{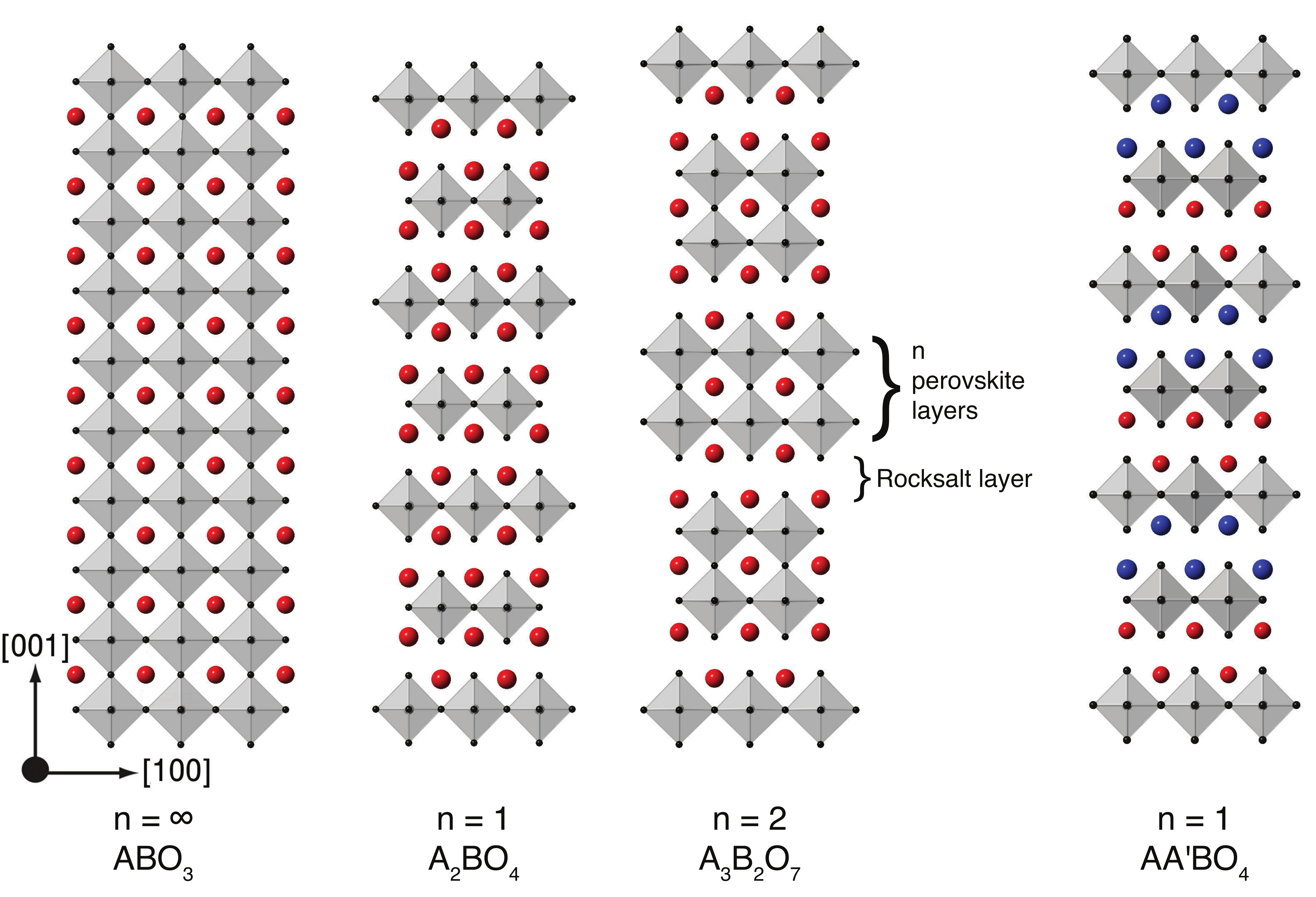}
\hspace{8mm}
\caption{The Ruddlesden-Popper A$_{n+1}$B$_n$O$_{3n+1}$ homologous series and an A-site ordered $n=1$ Ruddlesden-Popper phase (AA$^\prime$BO$_4$) built from two different A-site cations. The inversion center at the B-site is broken for the $n=2$ and A-site ordered AA$^\prime$BO$_4$ phases but is retained for the perovskite end member and the A$_2$BO$_4$ $n=1$ phase.}
\label{RPs}
\end{figure}

\begin{figure}
\centering
\includegraphics[height=11cm]{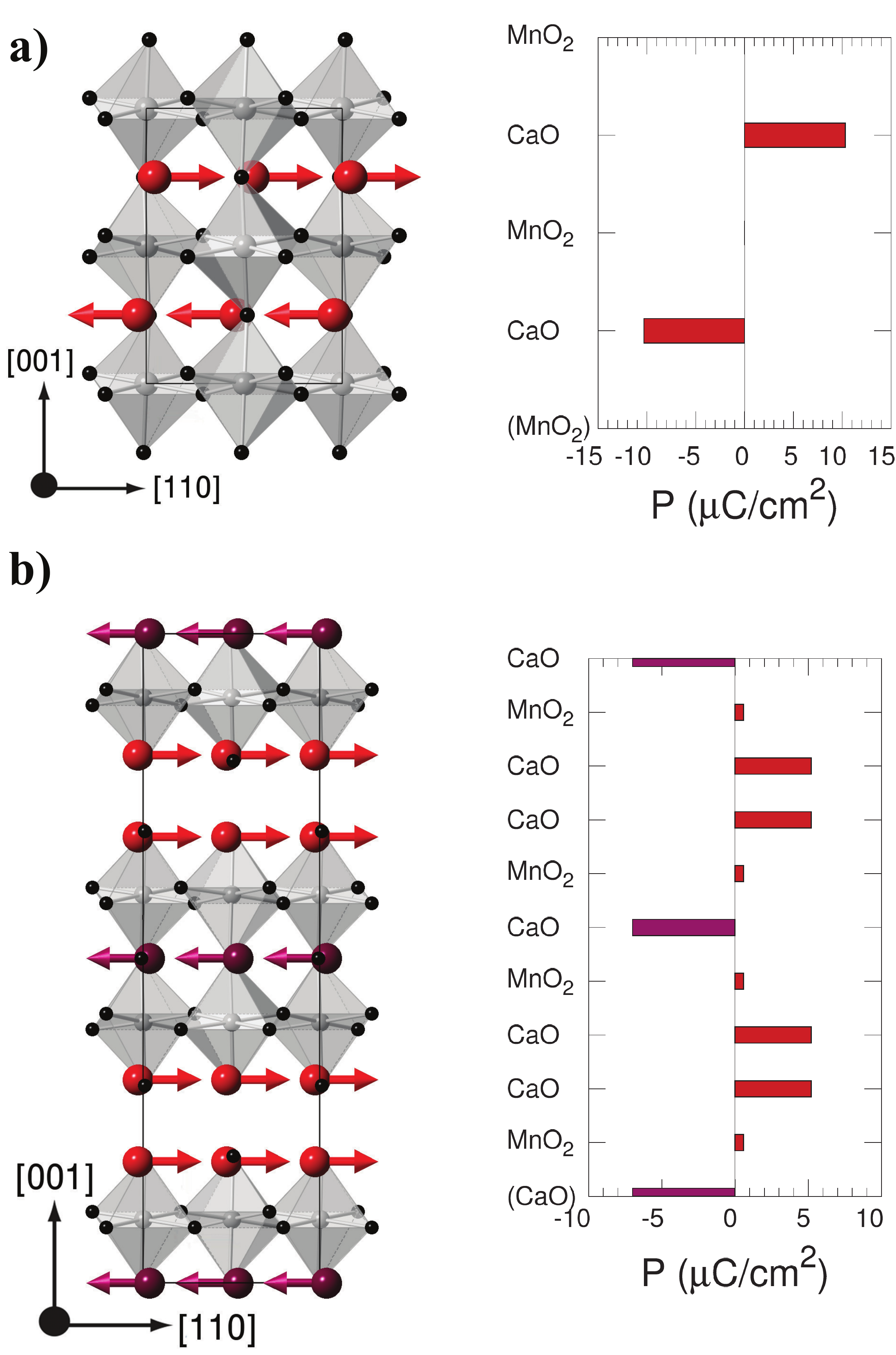}
\caption{The layer-resolved polarization of (a) CaMnO$_3$ in the \textit{Pnma} structure showing the antipolar displacements of the A-site cations and (b) The Ca$_3$Mn$_2$O$_7$ structure and layer-resolved polarization. This figure makes clear how layering CaMnO$_3$ to make an $n=2$ RP phase creates a net polarization from noncancellation of A-site cation displacements in inequivalent AO layers. The polarization is parallel to [110]. \textit{Methods: VASP, PAW-PBE potentials, PBEsol functional, 5$\times$5$\times$5 \textbf{k}-grid for CaMnO$_3$ and 4$\times$4$\times$4 \textbf{k}-grid for Ca$_3$Mn$_2$O$_7$, 600eV cutoff. Collinear G-type antiferromagnetism for both structures. U=4.5eV, J=1.0eV. Born charges calculated using DFPT in the Ca$_3$Mn$_2$O$_7$ ground state structure ($A2_1am$).}}
\label{P_CaMnO3}
\end{figure}

%
%
%
\subsubsection{Hybrid Improper Ferroelectricity in the Ruddlesden-Popper Phases}
One of the key lessons so far has been that the three-dimensional connectivity of the perovskite octahedra does not allow rotations to remove the inversion center at the B-site, so an additional kind of distortion or perturbation is required to allow hybrid improper ferroelectricity.
Many ABO$_3$ perovskite oxides naturally form layered structures (referred to as ``layered perovskites'') with two-dimensional connectivity of the octahedra. One example of a layered perovskite is the Ruddlesden-Popper (RP) phase, a homologous series of structures with general formula A$_{n+1}$B$_n$O$_{3n+1}$. Any given member of the RP series consists of ABO$_3$ perovskite blocks stacked along the [001] direction with an extra AO sheet inserted every $n$ perovskite unit cells  [Figure \ref{RPs}]. Hence, one can also write the chemical formula for the RP phases as (ABO$_3$)$_n$/(AO). We can think of the RP phases as one example of Nature's heterostructures. 

Given the discussion of the last section, we now consider whether a Ruddlesden-Popper phase formed from a $Pnma$ perovskite would be a hybrid improper ferroelectric.
Let us start with the $n=2$, A$_3$B$_2$O$_7$ structure. These materials are related to their perovskite building blocks through the insertion of an additional AO layer for every two perovskite unit cells in the [001] direction. The breaking of the octahedral connectivity along [001] removes the inversion center at each B-site. For one, this makes the AO layer between perovskite blocks symmetrically inequivalent to those within the rocksalt layers, that is, the AO layers immediately above and immediately below the BO$_2$ layer are symmetrically inequivalent (this is the chemical criterion of Rondinelli and Fennie).
Additionally, the disconnection of the octahedra by the extra AO layer leads to an odd number of AO layers within the perovskite block, so any $Pnma$ type of antipolar AO motion that induces a polarization will not exactly cancel, leading to a net polarization. It would thus seem that $n=2$ Ruddlesden-Popper phases of ABO$_3$ perovskites with $Pnma$ ground states may indeed be good candidates for hybrid improper ferroelectrics.
%
%

CaTiO$_3$ and CaMnO$_3$ are two $Pnma$ compounds in which several members of the RP series have been previously synthesized.
For the manganite, the $n=1,\;2$ and 3 members have been reported~\cite{fawcett98,martin01,lobanov04} whereas for the titanate, only the $n=2$ and 3 members have been synthesized and characterized so far~\cite{white91}. 
Curiously, it has been known for some time that the $n=2$ members of both materials (Ca$_3$Ti$_2$O$_7$ and Ca$_3$Mn$_2$O$_7$) display an octahedral rotation pattern similar to those of the parent perovskites but these Ruddlesden-Popper phases form in the nonpolar $A2_1am$ space group.
The possibility for rotation-driven ferroelectricity in these $n=2$ RP structures was anticipated some time ago from symmetry arguments~\cite{aleksandrov95}. However, it wasn't until recently that the origin of the polarization in these materials was elucidated. 

\begin{figure}
\centering
\includegraphics[height=7.5cm]{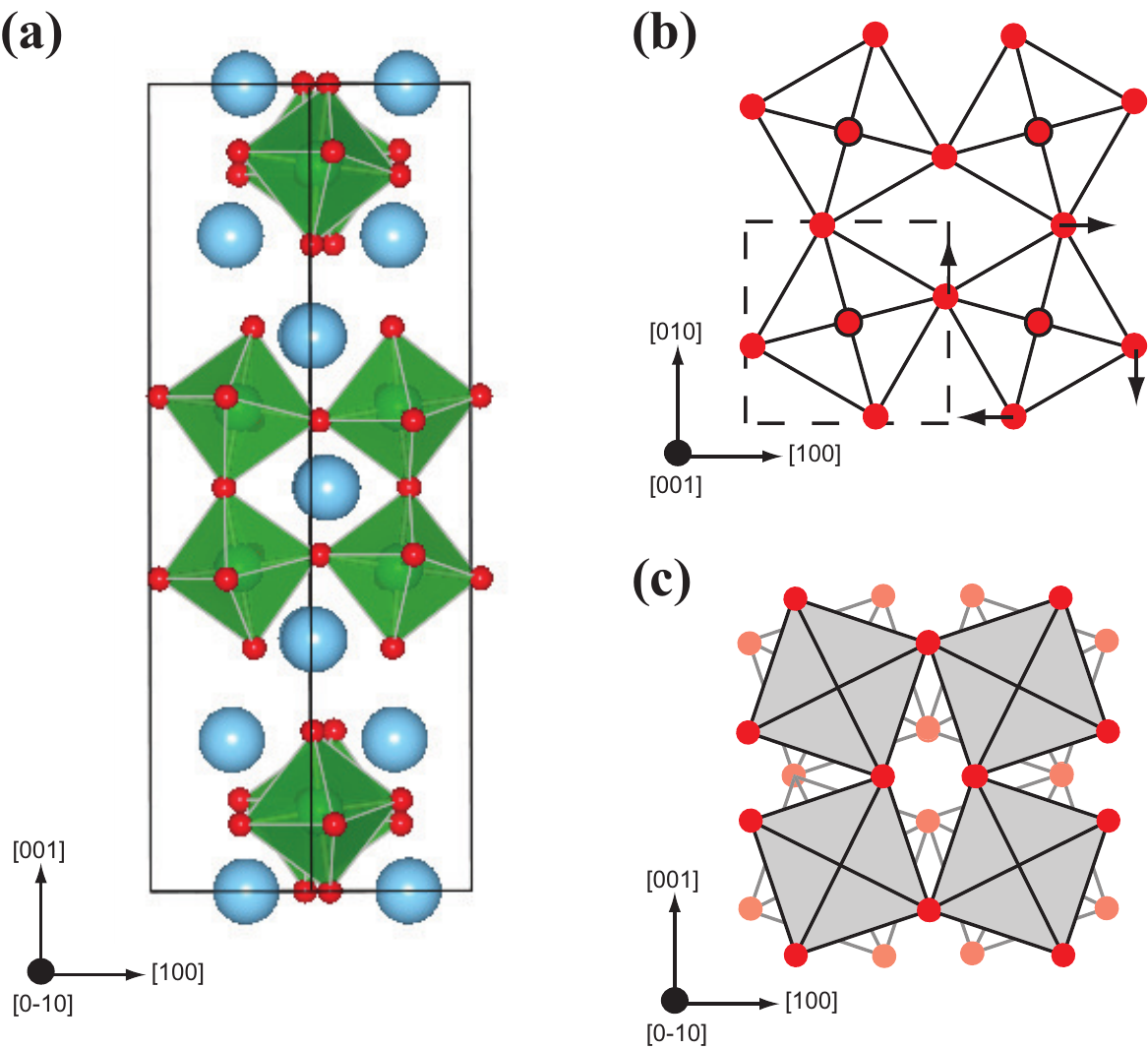}
\caption{(a) The $A2_1am$ ferroelectric ground state structure of Ca$_3$Ti$_2$O$_7$ and Ca$_3$Mn$_2$O$_7$. Large (blue) spheres correspond to Ca ions. (b) Schematic of the atomic displacements corresponding to the $X_2^+$ rotation mode. The dashed square denotes the unit cell of the $I4/mmm$ reference structure. (c) Schematic of the $X_3^-$ rotation mode. All axes refer to the coordinate system of the $I4/mmm$ reference structure. From Ref. 20.}
\label{RP}
\end{figure}

Using first-principles theoretical calculations, Benedek and Fennie~\cite{benedek11} investigated the lattice dynamical properties of Ca$_3$Ti$_2$O$_7$ and Ca$_3$Mn$_2$O$_7$. They showed that the transition from a paraelectric $I4/mmm$ reference structure to the ferroelectric $A2_1am$ ground state structure was driven by two different octahedral rotations with different symmetries. A polar instability is not involved. 
Ca$_3$Ti$_2$O$_7$ was calculated to have a large polarization of $P\:\approx\:20\:\mu$C/cm$^2$ and Ca$_3$Mn$_2$O$_7$ was found to have $P\:\approx\:5\:\mu$C/cm$^2$. Benedek and Fennie showed that the polarization arose from a similar trilinear coupling as in the SrTiO$_3$/PbTiO$_3$ superlattices (Equation \ref{trilinear}).  
Specifically, the single hybrid distortion pattern consists of two octahedral rotation modes: $Q_{X_2}$ ($R_1$), which is a rotation about the [001] axis  and transforms like the irrep $X_2^+$ (this rotation added to $I4/mmm$ leads to space group \textit{Cmca}) and the other, $Q_{X_3}$ ($R_2$), is a rotation about [110] and transforms like $X_3^-$ (leading to \textit{Cmcm}).  For a complete symmetry analysis see, Ref.~\cite{harris11}. 
The atomic displacements corresponding to each individual distortion, $Q_{X_2}$ and $Q_{X_3}$, are analogous to the a$^0$a$^0$c$^+$ and a$^-$a$^-$c$^0$ rotation patterns in the parent perovskites and are sketched in Figures \ref{RP}(b) and (c) respectively. Most importantly, as in the A-site ordered double perovskites, these rotation distortions induce displacements of the A-site cations in the AO layers, each of which generates a polarization. The topology of the Ruddlesden-Popper structure ensures a net non-zero polarization, as shown in Figure \ref{P_CaMnO3}(b). 

Other members of the Ruddlesden-Popper series can, in principle, also display octahedral rotation patterns that lead to nonpolar structures. In his symmetry analysis of octahedral rotation patterns in the Ruddlesden-Popper phases, Aleksandrov found that certain combinations of octahedral rotation patterns lead to polar structures in even $n$ members but nonpolar structures in odd $n$ members.\cite{aleksandrov95} We now understand that this is because the even $n$ members have two symmetrically inequivalent AO layers (hence the rotations induce A-site cation displacements that are not equal and opposite in each AO layer) whereas the odd $n$ members do not. In this sense, the odd $n$ members of the series are similar to the perovskite end member in that rotations by themselves do not satisfy the criteria for rotation-driven ferroelectricity.\cite{rondinelli11} However, it may still be possible to engineer rotation-driven ferroelectricity in the odd $n$ members by using the prescription presented for the perovskite end member: a layered arrangement of A/A$^\prime$ cations (see Figure \ref{RPs}). If each perovskite block in the RP phase separates chemically distinct double AO layers, a combination of two different octahedral rotations can lead to a nonpolar structure. Hence, hybrid improper ferroelectricity is not restricted only to the $n=2$ member of the Ruddlesden-Popper series.

%
\subsection{Other Examples of Rotation-Driven Ferroelectrics}  

Ferroelectricity induced by octahedral rotations is an exciting development in the field of ferroelectrics and although it is still being developed, the idea is gaining rapid interest.  There have been several other recent studies on materials in which octahedral rotations drive ferroelectric distortions. 

Fukushima, \textit{et al.,} showed that the main distortions contributing to the structure of the polar $P2_1$ phase of the double perovskite NaLaMnWO$_6$\cite{knapp06} are a pair of MnO$_6$ and WO$_6$ octahedral rotation modes.\cite{fukushima11}  This material is directly analogous to that of the PbTiO$_3$/SrTiO$_3$ superlattices in that a large polarization ($\sim 16\: \mu$C/cm$^2$) is induced through a trilinear coupling of a polar mode with the octahedral rotation modes. Unlike in Ca$_3$Mn$_2$O$_7$, where all the polar modes are stable, the polar mode in NaLaMnWO$_6$ is unstable but its amplitude in the ground-state structure is significantly enhanced through the coupling to the octahedral distortions. The authors surmise that it is the layered ordering of the Na and La ions that allows the usually nonpolar octahedral distortions to induce ferroelectricity.

Another material in which ostensibly nonpolar octahedral distortions induce ferroelectricity is the layered perovskite La$_2$Ti$_2$O$_7$. La$_2$Ti$_2$O$_7$ has one of the highest Curie temperatures (1770 K) of all known ferroelectrics and is a potential high-$T$ piezoelectric. Using first-principles calculations, L\'{o}pez-P\'{e}rez and \'{I}\~{n}iguez showed\cite{jorge11} that the ferroelectric instability in La$_2$Ti$_2$O$_7$ is an octahedral rotation mode similar in nature to those found in bulk ABO$_3$ perovskites. The octahedra in La$_2$Ti$_2$O$_7$ are only continuously connected in two dimensions. In the remaining dimension, La$_2$Ti$_2$O$_7$ has a layered structure where the perovskite blocks are separated by rocksalt layers. This is similar to the Ruddlesden-Popper compounds except the stacking is along the [110] direction of the bulk perovskite, whereas the RP stacking is along [001].  This difference in layering directions has the effect of allowing a single octahedral rotation to induce a spontaneous polarization in  La$_2$Ti$_2$O$_7$ (as opposed to the hybrid mode required in the Ruddlesden-Popper phases). The family of ferroelectric BaMF$_4$ fluorides,\cite{ederer06} where M = Mn, Fe, Co and Ni, are the $n=2$ members of the same family\cite{lichtenberg01} as the $n=4$ La$_2$Ti$_2$O$_7$.

%
%
\section{\label{coupling}Hybrid Improper Ferroelectricity and Strong Polarization-Magnetization Coupling}  
In the Introduction we briefly discussed the challenges involved in finding or creating cross-coupled multiferroics. Although we know that octahedral rotations in perovskites can significantly affect the magnetic properties of materials, rotations do not directly couple to electric fields in ABO$_3$ perovskites. Conversely, the type of structural distortion associated with ferroelectricity in prototypical ferroelectrics usually does not appreciably influence the interaction between spins\cite{wojdel09}.
Octahedral rotation-driven ferroelectricity solves these problems because the distortion that induces the polarization (octahedral rotations) also affects the magnetic properties. An advantage of the hybrid improper mechanism is that because the primary order parameter (the hybrid mode) consists of two rotation distortions of different symmetry, there are two independent lattice degrees of freedom that can potentially be exploited to control the magnetization.
In the previous section, we noted that the Ca$_3$Mn$_2$O$_7$ Ruddlesden-Popper phase is both a weak ferromagnet and a hybrid improper ferroelectric. Can the magnetization in this material be controlled with an electric field?

Benedek and Fennie showed, using a combination of symmetry arguments and first-principles calculations, that the $X^+_2$ rotation of Ca$_3$Mn$_2$O$_7$  induces a linear magnetoelectric effect; this electric-field tunable oxygen rotation distortion may lead to an enhanced magnetoelectric effect\cite{delaney09}. Additionally, the $X_3^-$ rotation mode induces weak ferromagnetism, that is, a small canting of the nominally antiferromagnetic spins, $\mathbf{S}_i$, through the Dzyaloshinskii-Moriya (DM) interaction\cite{dzy58,moriya60} or crystalline anisotropy.  Ca$_3$Mn$_2$O$_7$ is thus a multiferroic with strong polarization-magnetization coupling.

\begin{figure}
\centering
  \includegraphics[height=5.5cm]{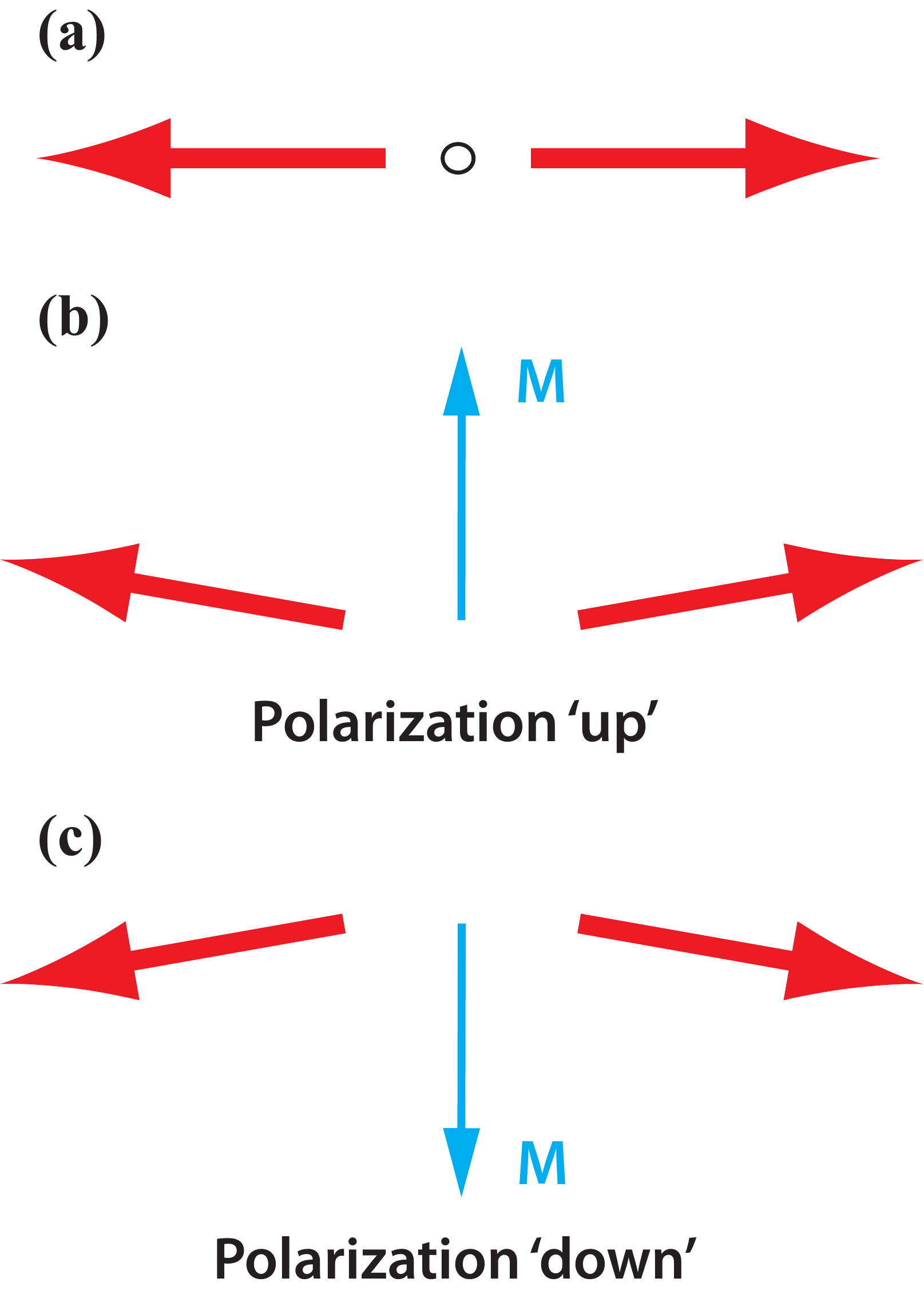}
  \caption{Schematic illustration of ferroelectrically-induced weak ferromagnetism through the DM interaction. (a) In the paraelectric antiferromagnetic state, the spins (depicted as red arrows) are related by inversion symmetry, $\mathbf{P} = 0$ and the Dzyaloshinskii vector $\mathbf{D} = 0$. (b) If a ferroelectric phase transition breaks inversion symmetry, $\mathbf{D}$ can become non-zero and induce a magnetization through spin canting. (c) An electric field can be used to switch $\mathbf{P}$, which would automatically change the sign of $\mathbf{D}$ and hence reverse the magnetization.}
  \label{DMfig}
\end{figure}

Unlike in a proper or conventional improper ferroelectric, more than one lattice distortion can switch the polarization in a hybrid improper ferroelectric. In the case of Ca$_3$Ti$_2$O$_7$ and Ca$_3$Mn$_2$O$_7$, the direction of the polarization can be switched by switching either the $X_2^+$ rotation mode, or the $X_3^-$ rotation mode, but not both. The issue of switching is particularly interesting in the case of Ca$_3$Mn$_2$O$_7$ because of the weak ferromagnetism\cite{lobanov04}. As previously discussed~\cite{scott77, fennie08,ederer08}, ferroelectrically-induced weak ferromagnetism\cite{varga09} is a mechanism -- the \textit{only known} mechanism -- by which an electric field can be used to switch the direction of the magnetization 180$^{\circ}$ in a single-phase material (see Fig.~\ref{DMfig}).  Since $X_3^-$ induces weak ferromagnetism, if $X_3^-$ switches when the polarization is reversed, the magnetization will also reverse.

The discussed switching experiment relies on two important assumptions. First, from symmetry arguments, the antiferromagnetic vector may change sign with $X_3^-$, instead of the magnetization. Switching of the canted ferromagnetic moment is usually more energetically favorable, but this does not have to be the case. Second, it is difficult to prove using first-principles calculations alone whether or not the polarization can be switched to a symmetry-equivalent state with an electric field small enough to be practically applied in a laboratory. Ca$_3$Mn$_2$O$_7$ does satisfy Abrahams' structural criteria for ferroelectricity\cite{abrahams68} but the switching experiment itself may be challenging.
Still, Ca$_3$Mn$_2$O$_7$ is a material where $R_1$ induces a linear magnetoelectric effect, $R_2$ induces weak ferromagnetism, and the combination of $R_1$ and $R_2$ induces ferroelectricity, all of which should lead to rich structural, (anti)ferromagnetic, and magnetoelectric domain configurations. Spatially resolving these individual domains -- optically, for example~\cite{fiebig} -- should prove the coupling physics discussed even without performing electrical switching experiments. We are sure that other materials like and superior to Ca$_3$Mn$_2$O$_7$ also await design and discovery.

%
%
\section{\label{end}Functional Octahedral Rotations}
One of the most exciting implications of the highlighted work is the prospect of \textit{functional} octahedral rotations: octahedral distortions that respond in a useful way to external perturbations, such as electric and magnetic fields. 
We have been discussing magnetic properties and multiferroics but one could pursue many other directions.
It is understood~\cite{hwang95} that the electronic properties of transition metal perovskites can be significantly altered by changing the octahedral rotation angle, a key ingredient  to realize such exotic electronic properties as colossal magnetoresistance. The hybrid improper mechanism opens the door to electric-field control of octahedral rotations and electronic properties.

Note that in materials that allow hybrid improper ferroelectric coupling, any two of the three distortions -- $P$, $R_1$, or $R_2$ -- making up the trilinear invariant may be the primary order parameters driving the system into the ground state. The novel idea presented by Bousquet and co-workers is for the possibility to achieve a type of rotation-driven ferroelectricity where the rotations $R_1$ or $R_2$ induce a polarization $P$ in the absence of a zone-center ferroelectric instability.
A question remains: when does a material that has a trilinear term allowed by symmetry behave more like a conventional improper ferroelectric and when does it behavior more like a proper ferroelectric? This is, in our view, the key question that needs to be explored in these new classes of materials because the ``ferroelectric'' properties will be different in these two limiting cases; this may be one of the reasons why several known materials have not been previously identified as being (hybrid improper) ferroelectrics.

We have described how octahedral rotations can form the basis of a mechanism for the electric-field switching of the magnetization in multiferroic materials. Our review has focused on bulk materials, but researchers are becoming increasingly interested in the properties of oxide interfaces and superlattices\cite{mannhart10}. It would be interesting to explore the possibility of using functional octahedral rotations to create functional interfaces: interfaces whose properties (electronic, magnetic, optical, catalytic) could be directly manipulated by electric or magnetic fields. The concept of functional octahedral rotations is not fundamentally limited to oxides either. Many different materials families feature anion polyhedra as a key structural unit and it is therefore likely that other versions of the hybrid improper mechanism described here exist in those material classes. The field of functional octahedral rotations is clearly ripe for discovery and exploration and we hope this contribution stimulates further progress in this area.

\section*{Acknowledgements}
We thank James Rondinelli, Max Stengel, Jorge $\acute{\rm I}$$\tilde{\rm n}$iguez, Patrick Woodward and Philippe Ghosez for useful discussions.
N.\ A.\ B.\ was supported by the Cornell Center for Materials Research with funding from the NSF MRSEC program (cooperative agreement no.\ DMR 0520404). A.\ T.\ M.\ was supported by NSERC of Canada and the NSF under grant no.\ DMR-1056441. C.\ J.\ F.\ was supported by the Department of Energy, Division of Basic Energy Sciences under grant no.\ DE-SCOO02334.

\bibliographystyle{model1a-num-names}
\bibliography{benedek}







\end{document}